\documentclass[twocolumn,tighten]{aastex631}

\hypersetup{colorlinks, citecolor=blue, linkcolor=blue}
\definecolor{fgreen}{rgb}{1,0.6,0.1} 
\newcommand{\trev}[1]{{{#1}}}

\newcommand{\mne}[1]{$-$}

\newcommand{\qdist}[1]{\ifmmode\langle#1\rangle\else\textlangle#1\textrangle\fi}
\newcommand{\mmu}{<\!\!\mu_{g}\!\!>}
\newcommand{\kms}{km s$^{-1}$}

\received{\today}
\revised{\today}
\accepted{\today}
\submitjournal{ApJ Supplement}

\shorttitle{Catalog of early-type dwarf galaxies}
\shortauthors{Paudel et al.}

\begin{document}

\title{An Extensive Catalog of Early-type Dwarf Galaxies in the Local Universe: Morphology and Environment}
\author[0000-0003-2922-6866]{Sanjaya Paudel}
\author[0000-0002-1842-4325]{Suk-Jin Yoon}
\affil{Department of Astronomy, Yonsei University, Seoul 03722, Republic of Korea}

\author[0000-0002-6841-8329]{Jaewon Yoo}
\affil{Korea Astronomy and Space Science Institute (KASI), 776 Daedeokdae-ro, Yuseong-gu, Daejeon 34055, Republic of Korea}

\author[0000-0001-5303-6830]{Rory Smith}
\affil{Korea Astronomy and Space Science Institute (KASI), 776 Daedeokdae-ro, Yuseong-gu, Daejeon 34055, Republic of Korea}

\author{Daya Nidhi Chhatkuli}
\affil{Central Department of Physics, Tribhuvan University, Kirtipur, Kathmandu, Nepal}

\author{Rajesh Kumar Bachan}
\affil{Department of Physics, Patan Multiple College, Tribhuvan University, Nepal}

\author{Binil Aryal}
\affil{Central Department of Physics, Tribhuvan University, Kirtipur, Kathmandu, Nepal}

\author{Binod Adhikari}
\affil{Department of Physics, St. Xavier's College, Tribhuvan University, Kathmandu, Nepal}
\affil{Department of Physics, Patan Multiple College, Tribhuvan University, Nepal}

\author{Namuna Adhikari}
\author{Amrit Sedain}
\author{Sharup Sheikh}
\author{Sarashwati Dhital}
\affil{Central Department of Physics, Tribhuvan University, Kirtipur, Kathmandu, Nepal}

\author{Ashutosh Giri}
\affil{Department of Physics, St. Xavier's College, Tribhuvan University, Kathmandu, Nepal}
\author{Rabin Baral}
\affil{Department of Physics, St. Xavier's College, Tribhuvan University, Kathmandu, Nepal}


\correspondingauthor{Suk-Jin Yoon}
\email{sjyoon0691@yonsei.ac.kr }

\begin{abstract}
We present an extensive catalog of 5405 early-type dwarf (dE) galaxies located in the various environments, i.e., clusters, groups and fields, of the local universe ($z$ $<$ 0.01).
The dEs are selected through visual inspection of the Legacy survey's $g$-$r$-$z$ combined tri-color images. 
The inspected area, covering a total sky area of 7643 deg$^{2}$, encompasses two local clusters, Virgo and Fornax, 265 groups, and the regions around 586 field galaxies of $M_{K}$ $<$ $-$21 mag.
The catalog aims to be one of the most extensive and publicly accessible collections of data on dE, despite its complex completeness limits that may not accurately represent its statistical completeness.
The strength of the catalog lies in the morphological characteristics, including nucleated, tidal, and ultradiffuse dE. 
The two clusters contribute nearly half (2437 out of 5405) dEs, and the 265 groups contribute 2103 dEs. 
There are 864 dEs in 586 fields, i.e., $\sim$\,1.47 dEs per field. 
Using a standard definition commonly used in literature, we identify 100 ultra-diffuse galaxies (UDGs), which take $\sim$\,2\,\% of the dE population. 
We find that 40\,\% of our sample dEs harbor a central nucleus, and among the UDG population, a majority, 79\%, are nonnucleated. About 1.3\,\% of dEs suffer from ongoing tidal disturbance by nearby massive galaxies, and only 0.03\,\% show the sign of recent dwarf-dwarf mergers. 
The association between dEs and their nearest bright neighbor galaxies suggests that dEs are more likely created where their neighbors are non-star-forming ones. 
\end{abstract}


\section{Introduction}
Stated as a morphology--density relation, galaxy population depends on the local environment \citep[e.g.][]{Dressler80,Whitmore93}. Massive galaxy clusters are dominated by red and dead early-type morphology galaxies, and low-dense groups and fields host a larger fraction of star-forming late-type galaxies \citep{Oemler74}. In the dwarf galaxy regime (commonly defined as less massive than the Large Magellanic Cloud (LMC)), we can see an extreme form of the morphology--density relation, with the non-existence of early-type morphology, dwarfs outside the group and cluster environments \citep{Binggeli87,Geha12}.  

A natural explanation for the existence of an extreme form of the morphology-density relation in the dwarf galaxy population is that the environment-related mechanisms mainly cause the quenching of dwarf galaxies and there are varieties of candidates \citep{Boselli08,Kormendy09}. Ram pressure stripping (RPS; high-speed interaction between hot cluster gas and reservoir of star-forming gas of in falling galaxy introduces a viscous drag, which eventually removes the star-forming gas \citep{Gunn72}), tidal harassment/stripping (removal of gas through tidal deformations \citep{Moore96,Mayer01,Smith10}), and starvation (a short supply of star-forming gas in the cluster environment \citep{Larson80}), are the prime candidates that are frequently discussed in the literature. Since the environments themselves vary from the cluster and group to the field, a longstanding and critical issue is which particular mechanism is the more efficient in which particular environment. The RPS is expected to be more efficient in the cluster environment. Evidence of ongoing RPS in the cluster environment, explicitly in the Virgo cluster, have been frequently presented in the recent literature \citep{Kenney04,Kenney14,Vollmer01}. The tidal stirring or harassment is more frequent in the small group environments \citep{Mayer01,Paudel14a}. It is also important to note that the combination of time spent while undergoing a given mechanism as well as its strength, should be considered. It has been suggested that only a combination of several different processes can explain the diverse morphological properties of cluster early-type dwarf galaxies \citep[dEs, see e.g.,][for an overview]{Lisker09}. On the other hand, the importance of the environment relative to a small-scale merger in the evolution of dwarf galaxies remains poorly understood.

\begin{figure*}
\label{main}
\includegraphics[width=18cm]{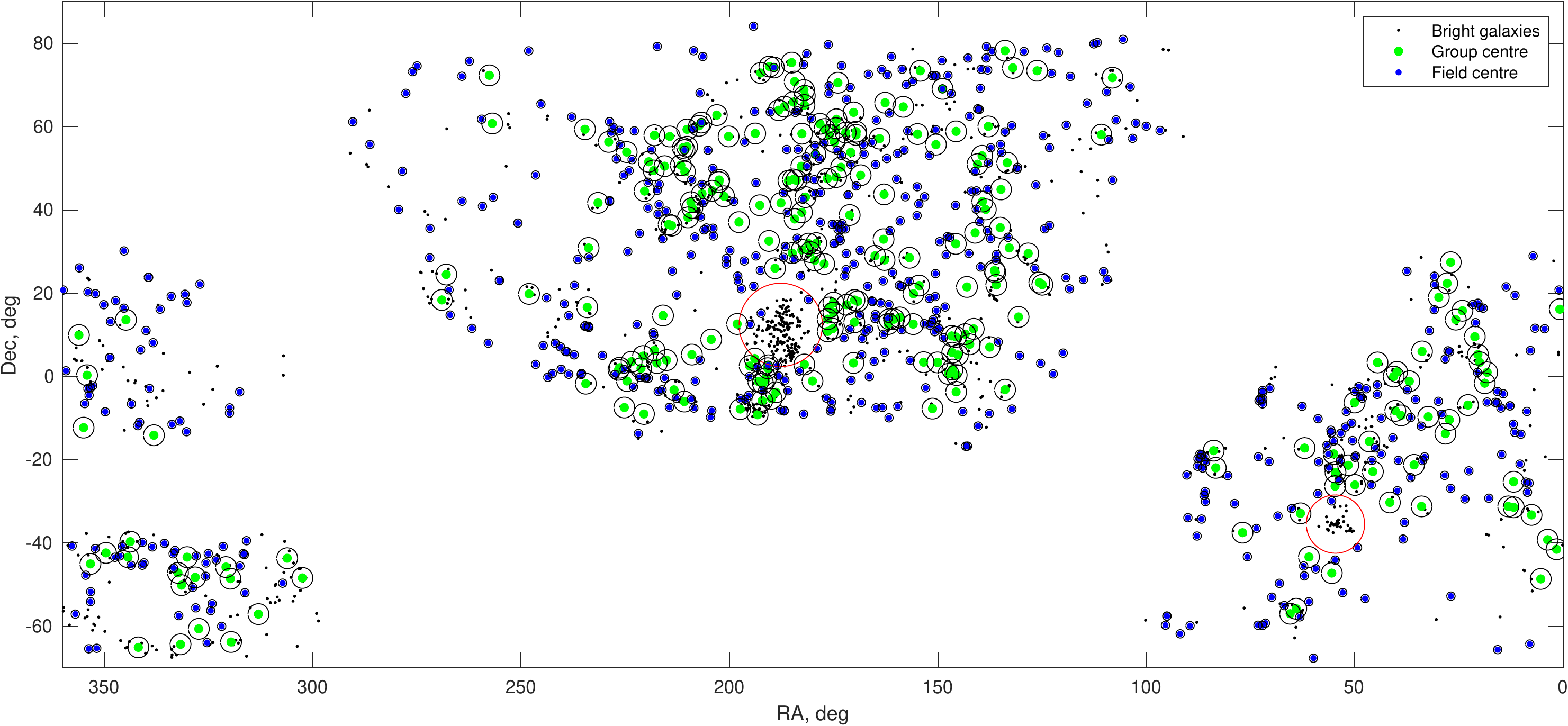}
\caption{All-sky map of our visually inspected area. We show the visually inspected areas of groups (shown in green) and fields (shown in blue) with circles of radius 2.5 and 1.0 degrees, respectively. We also mark the Virgo and Fornax regions by the red circle of radius 15 and 10 degrees, respectively. The small black dots represent bright (M$_{K}$ $<$ $-$21 mag) galaxies at z $<$ 0.01.}
\label{samplearea}
\end{figure*}

By definition, dEs are non-star forming and gas-poor objects, somewhat lower mass cousins of Elliptical galaxies \citep{Ferguson94}. Therefore, a smooth appearance and red and dead (aka old stellar populations) are the prime characteristics of these galaxies, but more notably, their low-surface brightness nature makes them an outstanding class of object compared with massive early-type galaxies. Due to their low-surface brightness and scarcity in the nearby universe, particularly in low-dense environments like the field, a detailed study of environmental effects on their observational properties using a large sample of dEs has been lacking. 

On the other hand, detailed studies on the morphological properties of cluster dEs have revealed that dEs are not a homogeneous class of objects. With the help of sophisticated image analysis techniques, a great deal of complexity in dEs structure has been discovered. Fine structures such as tidal tails, blue cores, spiral-arms or bars are frequent in the brighter dE population \citep{Jerjen00,Lisker07,Lisker06a,Paudel14a} and in general, cluster dEs light profiles are not well fitted by a single S\'ersic function \citep{Janz14,Su21}. A careful analysis of these fine structures has shown that they could contribute up to 10 percent of the total light of the dE main body \citep{Smith21,Paudel14a}. 

Very recent deep imaging surveys have also revealed that the dEs of stellar mass as low as the Fornax dwarf galaxy is hosting tidal features that are likely to have originated through mergers \citep{Crnojevic14,Paudel17a}.

In addition, a compact star cluster is also found in the center of many dEs, called the nucleus. Many detailed observational studies have shown that the central nucleus is almost universal in the brighter dE population of the cluster environment \citep{Cote06} and the nucleated fraction declines to as low as 30\% among the fainter ($M_{g}$ $>$ $-$12 mag) dEs \citep{Sanchez19}. Interestingly, a few dedicated studies on these sub-structural properties of dEs beyond the cluster environment, particularly outside of the Virgo and Fornax cluster, are available to date and it is not well understood that the cluster environment is necessary for forming these sub-structural properties.

There is a recent upsurge in the exploration of a diffuse and extended dE on the name of Ultra Diffuse galaxies \citep[UDGs,][]{Dokkum15,Marleau21}, although their examples have been known for decades \citep{Thompson93,Caldwell87,Impey88,Jerjen00,Conselice02,Conselice03,Penny11}. A multitude of recent literature has presented the UDGs,  as if a newly discovered class of dwarf galaxies \citep{Mihos15,Delgado16,Merritt16,Roman17,Zaritsky21} and there are reports of discovering them in all environments -cluster, group, and field, with a diverse color, morphology, and globular cluster population \citep{Beasley16,Lim20,Muller20}. The diverse and complex UDG's properties essentially suggest that UGDs could also form a mixed bag of populations, and there is no single origin for them, with multiple possible evolutionary tracks at play \citep{Duc14,Amorisco16,Bennet18,Chan18,Carleton19,Jones21}.

With the advent of sophisticated observing techniques and wide-aperture telescopes, there has been a growing interest in low-surface brightness galaxies like dEs. Some groups have focused on building catalogs of dwarf galaxies with different morphology, and others have focused on a more specific type, like dEs or UDGs in the cluster or group environment. For example, \cite{Habas20} investigated the overall dwarf galaxy population of different morphologies around massive early-type galaxies, and \cite{Ferrarese12,Venhola17,Eigenthaler18,Ferrarese20,Marca22} are more focused on the cluster environment. As a part of the Satellites Around Galactic Analogs (SAGA) survey \cite{Geha12} searched dwarf galaxy populations around  Milky Way analogs and \cite{Carlsten22} explored satellite populations in local volume ($D$ $<$ 12 Mpc) host galaxies. In addition, \cite{Chiboucas13,Crnojevic16,Muller17,Bennet19} searched member dwarf satellite galaxies in nearby groups.

In this paper, we investigate the morphological properties of dEs sampled from diverse environments trying to understand their relationship with the environment where they are located and provide statistically homogeneous data sets of a large sample of dEs and their morphological properties. This paper is organized as follows: Section \ref{iden} briefly describes the dE candidate identification procedure. In section \ref{anal}, we present a measurement of photometric properties. Section \ref{morph} presents the results analysis of photometric and morphological properties. Finally, How our dEs are distributed in the cluster and group environment and their association with massive galaxies is described in section \ref{env}. Our main findings are concluded in section \ref{conc}.

\section{Identification}\label{iden}

A primary motivation of this work is to create a large sample of dEs located in various environments, i.e., cluster, group, and field, of the local universe $z$ $<$ 0.01. We first used the publicly available imaging and spectroscopic data of the Sloan Digital Sky Survey \citep[SDSS][]{Aihara11} to identify pure red and dead dwarf galaxies, selecting all dwarf galaxies of $M_{g}$ $ >$ -18 mag and visually classified them as dE. We classify a galaxy as dE if the SDSS color image shows a red (color index $g-r$ $>$ 0.5 mag; \cite{Lisker06}) and smooth appearance and the presence of no significant H$_{\alpha}$ emission in the SDSS optical spectrum. It is customary to use the first criteria to classify galaxies as dE in case the optical spectrum is not available. Since the SDSS also provides the optical spectrum, we have added the second constraint to confirm that we are selecting purely non-star-forming galaxies. 

\begin{table}
\includegraphics[width=8.5cm]{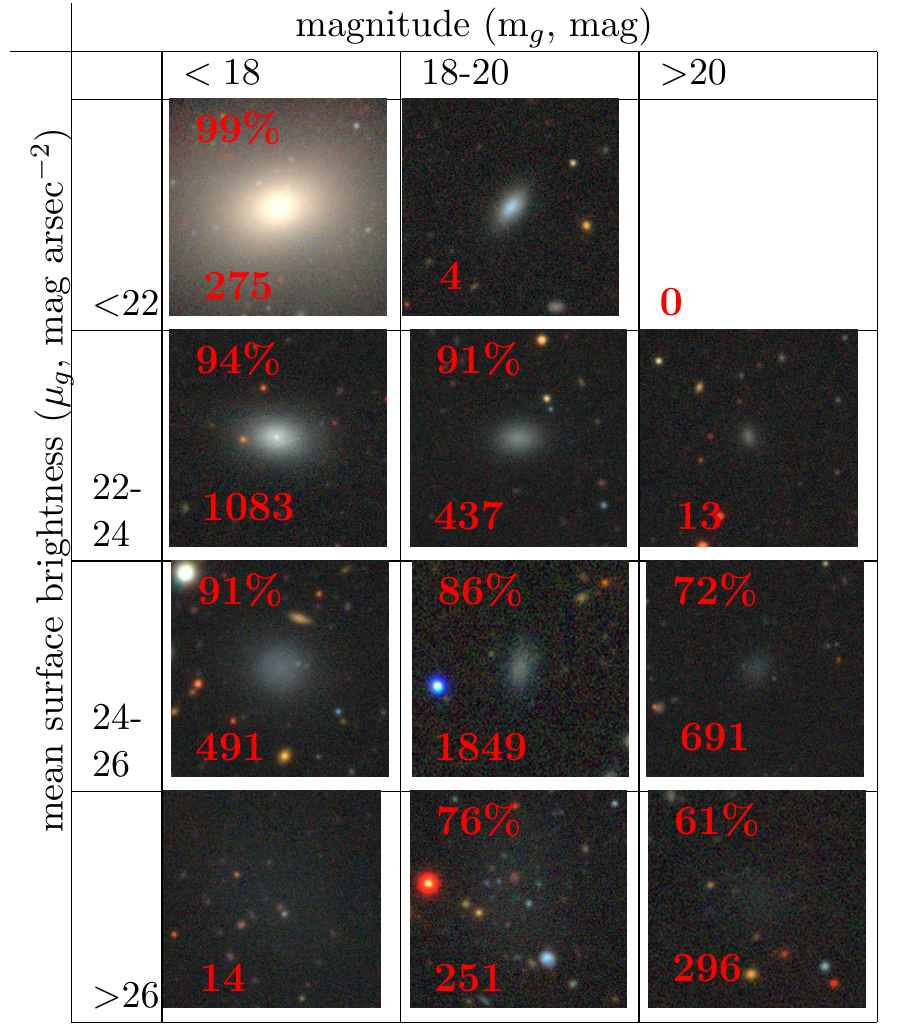}
\caption{Recovery fraction as a function of magnitude and mean surface brightness. The row represents the $\mmu$ bins and the column represents $g$-band magnitude bins. We show an example dE of each particular bin, randomly drawn from the sample. At each bin, the recovery fraction and the total number of dEs are denoted at the top and bottom, respectively. \trev{This analysis provides a representative sample of the recovery fraction but does not fully reflect the statistical completeness of the study.}}
\label{rcfrac}
\end{table}

Within the area covered by the SDSS, there are 786 dEs in the redshift range of less than 0.01. These spectroscopically selected dEs served as a training sample to further extend the dEs sample from a wider sky coverage imaging database provided by the Legacy survey \citep{Dey19}. This training is crucial, particularly for students who are starting to learn galaxy evolution research. 

Since the Legacy survey does not have spectroscopic observation, we only used visual inspection to identify the dEs using $g-r-z$ tri-color images provided by the Legacy survey visual tool. 

Contrary to the common practices, which first determine the sources/objects from a pre-built catalog that is primarily produced by an algorithm and then select the object of interest using specific criteria, we identified the dEs directly in the field, taking advantage of experienced knowledge of visual analysis of dE selected from the SDSS. Our method avoids the pre-selection of the sources detected by some automated software like source extractor \citep{Bertin96}. Given the heterogeneous morphology of dEs, including extremely diffuse UDG, the automated source detection may miss the faint low-surface brightness objects due to their low signal. Even detected, specific criteria may not cover all dEs-like objects.

To minimize the contamination from the compact, high-surface brightness interlopers (such as foreground stars, globular clusters, and high-redshift galaxies), we explicitly select low-surface brightness and extended galaxies, the prime characteristics of dE. This selection may be biased against the compact early-type dwarf galaxies, such as M32. Nevertheless, they are several rare kinds of objects, e.g., there are a hand-full of cE like compact non-star forming dwarf galaxies while there is more than 500 normal dE in the Virgo Cluster Catalog (VCC) of the Virgo cluster \citep{Janz08,Chilingarian09}.

To select a variety of environments that host dE, we have chosen all three essential aspects, i.e., cluster, group, and field. For the cluster, we have chosen two local clusters, Virgo and Fornax, and for the group, we have selected groups from a group catalog provided by \citep{Makarov11}. For the field, we define an area around bright galaxies of magnitude $M_{k}$ $<$ -21 mag, which are not a member of any cluster or group according to \cite{Makarov11} group catalog. We select the host environment to avoid distance uncertainty of newly detected dE. Since, by design, most newly identified will not have radial velocity information, we assume they are located at line-of-sight distances of the hosts. Note that the definition used here for the field dE could be inconsistent with that commonly used in the literature. Since we deliberately searched dEs around giant galaxies, they are indeed members of the satellite system of that host galaxy, assuming that their distances are similar to the host. The selected environment represents the environment of host galaxies, not those detected dEs. However, we use the terminology ``field dE" for those detected around the field host to simplify environment classification.

 In total, we selected two clusters, 265 groups and 586 fields within a redshift range between  $-$0.0013   and $<$ 0.01.  
 
Indeed, there is a large body of dE surveys in cluster environments, particularly for Virgo and Fornax \citep{Binggeli85,Venhola17,Ordene18,Ferrarese20}, which can be found in the literature. However, this study aims to provide a more homogeneous sampling of dEs in diverse environments ranging from cluster to field.

Finally, we visually inspected a 1 Mpc area around the group center, typically a 2.5-degree sky-projected radius for a group located at a distance of 25 Mpc. This distance corresponds to the median distance of our selected groups. We inspected a 15-degree radius around the Virgo and a 10-degree radius around the Fornax cluster. We inspected a 450 kpc area from the host center for the field; this corresponds to a one-degree angular radius for the field located at 25 Mpc, which is the median distance of our field sample.  

The motivation for selecting areas around the cluster, group, and area around bright Field galaxies is the finding of \cite{Geha12}, there are none or rare existence of non-star forming dwarf outside of group or cluster environment. In addition, from our spectroscopy sample, we noticed that very few dEs were located beyond  500 kpc of giant galaxies with a relative velocity larger than 500 \kms.

 We show an all-sky view of the inspected area covered by the Legacy survey DR9\footnote{https://www.legacysurvey.org/dr9}  in Figure \ref{samplearea}, where each circle represents a group area with a 2.5$^{\arcdeg}$ radius. The two nearby clusters, Virgo and Fornax, are defined by the red circle of 10 and 7-degree radii, respectively. As we can see, the inspected areas frequently overlap with one another; in total, we have inspected 7643 square degrees of the Legacy survey images.

For visual inspection purposes, we extensively use tri-color cutout images provided by sky-viewer of the Legacy survey\footnote{https://www.legacysurvey.org/viewer}. By design, we will not have radial velocity information for our dEs. Instead, based on their morphology and surface brightness, we assessed their heliocentric distance, less than 40 Mpc for $z$ = 0.01. The basic idea is our dEs have lower surface brightness than background sources for a given apparent magnitude. Furthermore, morphologically our dEs are smooth and regular and they follow a well-defined relation between absolute magnitude versus radius and surface brightness, which can also be used to constrain their heliocentric distances.

\begin{figure}
\label{main}
\includegraphics[width=8.5cm]{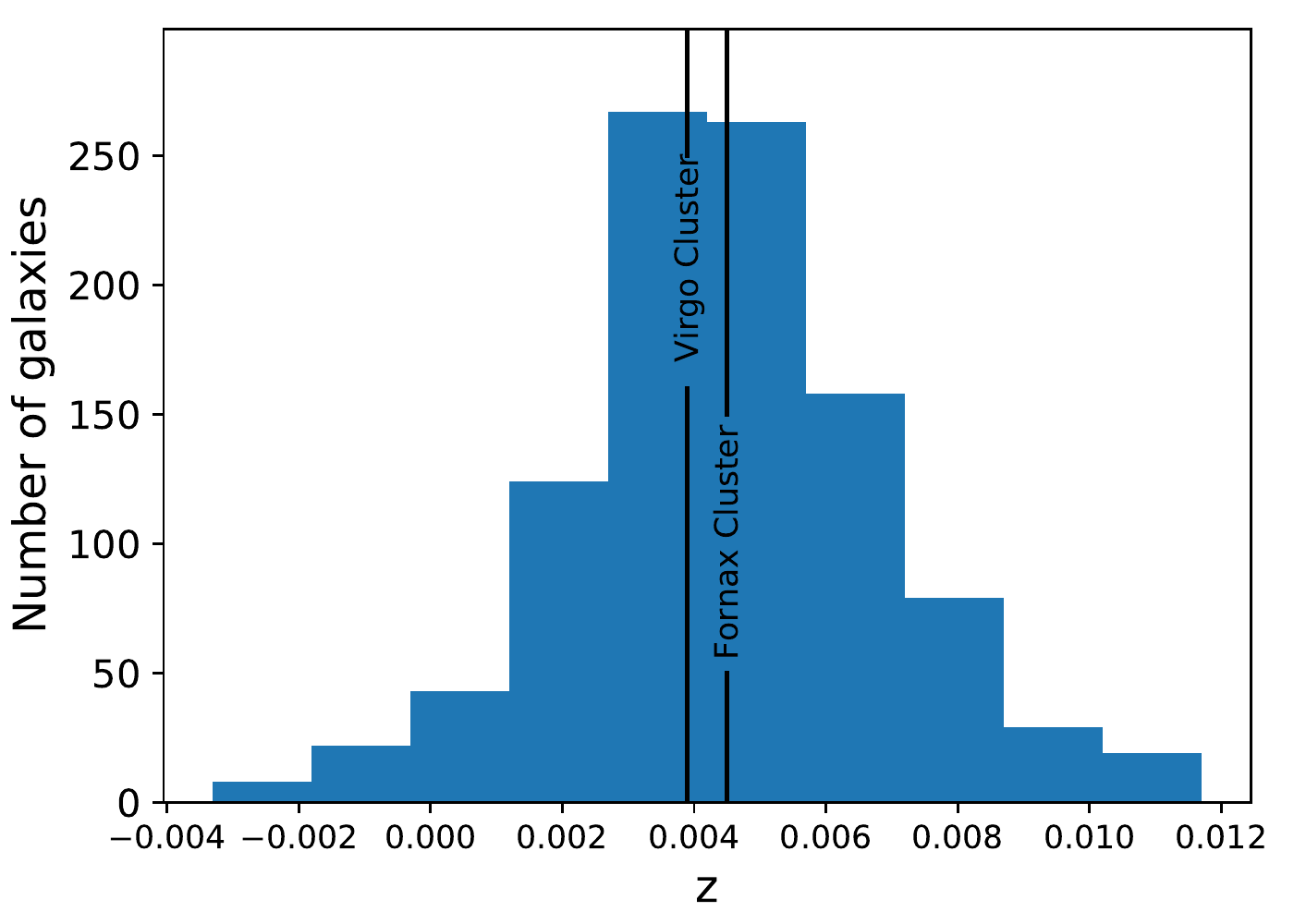}
\caption{Redshift distribution of dEs, for which radial velocity information is available. We also highlight the position of two clusters, Virgo and Fornax, by the vertical lines.}
\label{zhist}
\end{figure}

\begin{figure*}
\label{main}
\includegraphics[width=18cm]{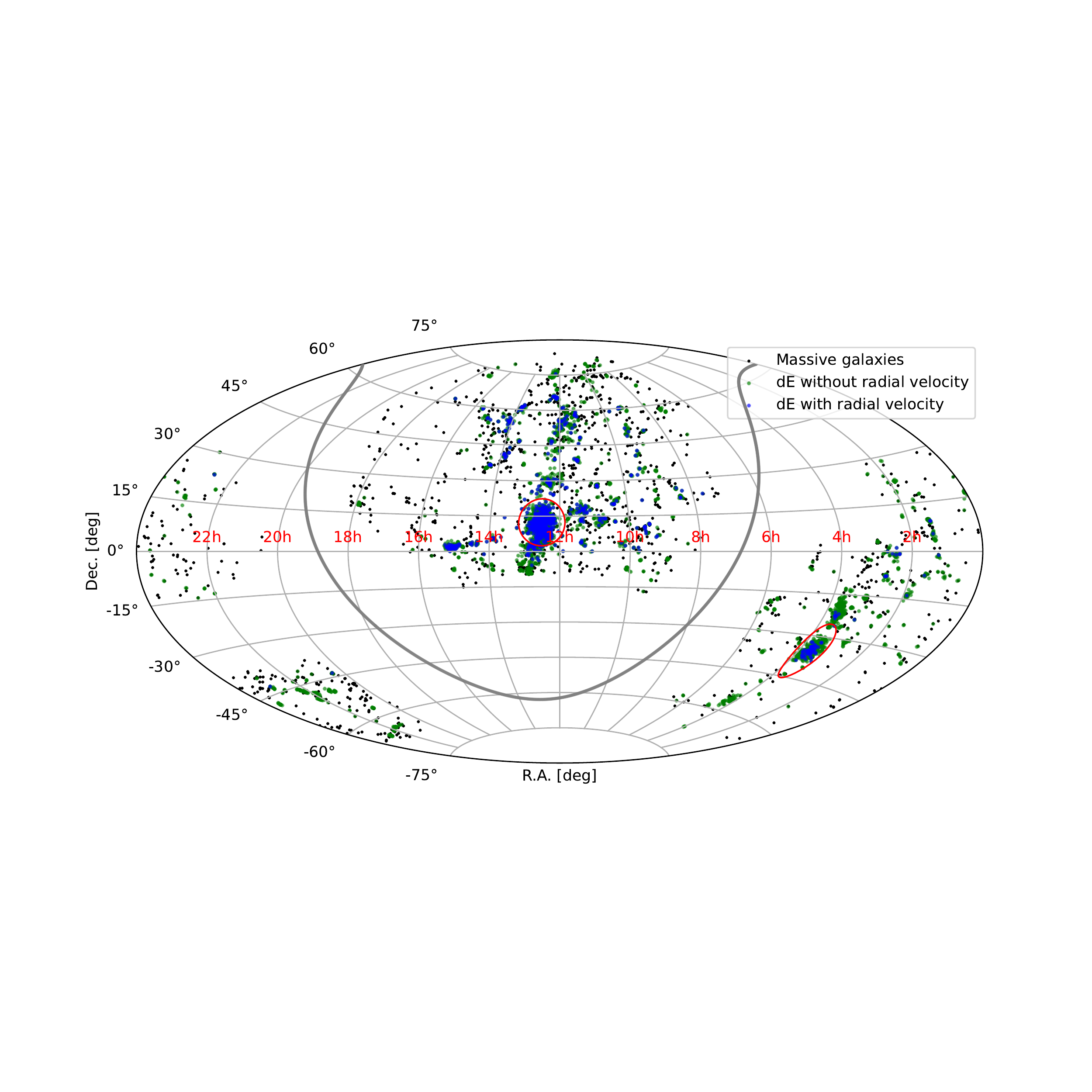}
\caption{All sky distribution of identified dEs in our sample. The sky coordinate is in the Atioff projection. The blue and green dots represent the sky position of dEs with and without radial velocity information, respectively. The black dots represent the sky positions of giant galaxies. We also highlight two cluster regions, Virgo and Fornax, using the red circles of 10-degree and 7-degree radii, respectively.}
\label{alsky}
\end{figure*}

The visual catalog is created by a simple visual inspection of tri-color cutout images provided by the sky-viewer. At least two persons visually check each field, and we find that there are quite good agreements between each other ($>85\%$).  Indeed such a visual scanning procedure is not 100\% complete, and there remains the possibility that some dEs in the field of view has been missed. In an effort to quantify the missing frequency, we rescanned the field after placing the previously identified dE in a randomly chosen position. The mock dEs were selected based on binning of magnitude and mean surface brightness, and we repeated this exercise 100 times for each designated bin. In Table \ref{rcfrac}, we list the recovery fraction of dEs in our identification method, where the results are binned with the width of 2 mags in both surface brightness (y-axis) and magnitude (x-axis). We also list the total number of dEs in each bin and note that we ran our experiment only for those bins that have more than 100 dEs. As we expected,  the recovery fraction depends on the magnitude and the mean surface brightness. We find that the recovery fraction is 99 percent for the brightest dE sample and decreases to 61 percent for the faintest sample. The median $g$-band magnitude of our sample dE is 19 mag and the magnitude range 18$-$20 covers two-thirds of our dEs sample (see section \ref{anal}); thus, we can consider that our recovery fraction is over 80\% on average. Our detection might also be influenced by the fact that we have used the SDSS confirmed dE (mostly brighter one) for the training purposes, and that may lead to a memory bias toward these bright dEs compared to the faint dEs, which is seen for the first time by the inspector.

An all-sky distribution of our dE sample is shown in Figure \ref{alsky}. To assign a host for a dE, we used the minimum radial separation between the host center and the dE; this is particularly important for the overlapping groups. For the Virgo Cluster member dEs, we used a circle of a 10-degree radius centering on M87 and for the Fornax Cluster, we used a circle of a 7-degree radius centering on NGC 3407. As we can see clearly, a large majority of dEs are concentrated around the clusters and large groups. The two clusters alone contribute nearly half (2437 out of 5405) of the total dE sample, and 265 groups contribute 2103 dEs. We have found 864 dEs in 586 fields. 

We have identified a total of 5405 dEs in the clusters, groups and fields; among them, 1324 dEs have radial velocity information (hereafter spectroscopic sample) obtained from the NED and the SDSS. We show the redshift distribution of the spectroscopic sample in Figure \ref{zhist}, and the peak of the distribution matched with the mean radial velocity of the Fornax and Virgo Clusters, and indeed, they dominate the spectroscopic sample dEs. The two clusters constitute more than two-thirds of spectroscopy sample dEs.

We list this main sample of dEs in Table \ref{catlog}. We consider distance to the dE similar to their host cluster, group or field. We consider 16.5 and 17.5 Mpc distance for the Virgo and Fornax Clusters and for the group we obtained it from \cite{Makarov11}. For the field galaxies that do not belong to any group or cluster we use Hubble flow distance, using  a standard cosmology with $H_{0}$ = 71 km s$^{-1}$ Mpc$^{-1}$, $\Omega$$_{m}$ = 0.3, and $\Omega$$_{\Lambda}$ = 0.7.

\section{Image Analysis}\label{anal}

Once we finalized the source catalog, we performed the image analysis and measured the photometric and structural parameters. For this purpose, we exclusively used the $g$-band stacked images provided by the Legacy survey.  We retrieved stacked images from the Legacy survey archive. While the Legacy image processing pipeline is optimized for relatively small objects, the photometric measurement of the large extended object is often inaccurate owing to imperfect sky subtraction and deblending of multiple sources. Galaxies with large angular sizes and irregular morphologies are often shredded into multiple separate objects, and thus the total derived fluxes for these galaxies are often unreliable. Furthermore, the Legacy survey photometric pipeline systematically underestimates the luminosities of galaxies of large sizes due to the overestimation of the sky background. In order to overcome these problems, we performed photometric measurements for all dE.  We retrieved 3\arcmin$\times$3\arcmin\, fits image cutout. We subtracted the sky background preparing a background map for each object. The background maps were constructed after masking out all identified sources in the image, which were defined by source-extractor segmentation maps, and  the segmentation images were filled by the median values calculated from surrounding all pixels. This method allows us to eliminate any contribution of light from stars and background galaxies. Finally, the background map was subtracted from the original fits file to remove the sky background contribution to the observed flux. 

\begin{figure}
\label{main}
\includegraphics[width=8cm]{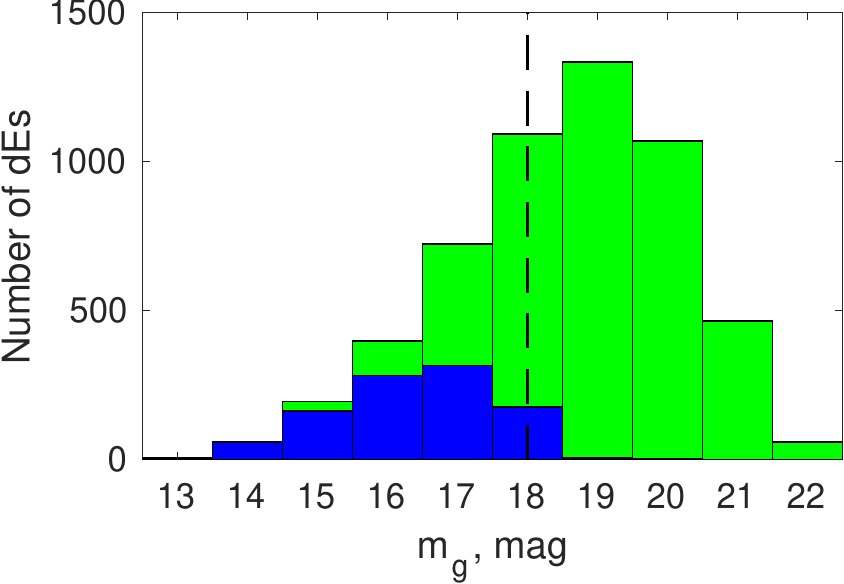}
\caption{The $g$-band magnitude distributions of our dE sample (green) and a subset having radial velocity information (blue). The vertical dashed line represents the magnitude limit of the spectroscopic target selection of the SDSS.}
\label{maghist}
\end{figure}

Further image analysis was done largely in the same way as in \cite{Paudel15}. All unrelated foreground and background objects were masked, and the centers of galaxies were calculated as the centroid of the flux distribution in the masked image. The total magnitudes were calculated using the Petrosian method \citep{Petrosian76}, for this, we derived an azimuthally averaged light profile using a circular aperture. The Petrosian radius is defined as the galactocentric radial position where the ratio of surface brightness at the radius $R$ to the average surface brightness within the $R$ reaches a certain value, denoted by $n$ i.e., \\

\begin{equation}
n(R) = \frac{\mu(R)}{\qdist{\mu(R)}}
\end{equation}

where $\mu(R)$ is surface brightness at radius $R$ , and $\qdist{\mu(R)}$ is the average surface brightness. We determine the Petrosian radii, $R_{p,n=2}$  and corresponding magnitudes, $m_{p,n=0.2}$, by adopting the most commonly used parameter $n_{R} = 0.2.$

We finally calculated the total brightness (m$_{g}$), summing up all pixel flux within the two $a_{p}$.  The half-light radius is then calculated where the total flux becomes half and uncommon to the usual practice our derived half-light radius is circularized, and it is not necessary to correct it by multiplying a scaling factor$ (b/a)^{0.5}$.

If the Petrosian radius did not converge, due to the fact that these galaxies sit within the light of nearby bright sources, or that are highly deblended/overlapped with the nearby foreground/background bright object, we performed manual aperture photometry in $g$-band  images only, after manually masking those unrelated objects. In that case, we will not have structural parameters derived for these galaxies.

The mean surface brightness ($\mmu$) was calculated using an equation \\

\begin{equation}
\qdist{\mu}_{1/2}=m+2.5log(2\pi R_{1/2}^{2})
\end{equation}

The derived magnitudes were corrected for Galactic extinction using \cite{Schlafly11}, but not $K$-correction. As expected, however, the $K$-corrections factors are negligible for the distance range of our target galaxies. Therefore, no $K$-correction has been performed in the derived magnitudes. Using the $g$-band as a reference, we followed the same procedure to measure photometric parameters in the $r$-band images.

\begin{figure}
\label{main}
\includegraphics[width=8cm]{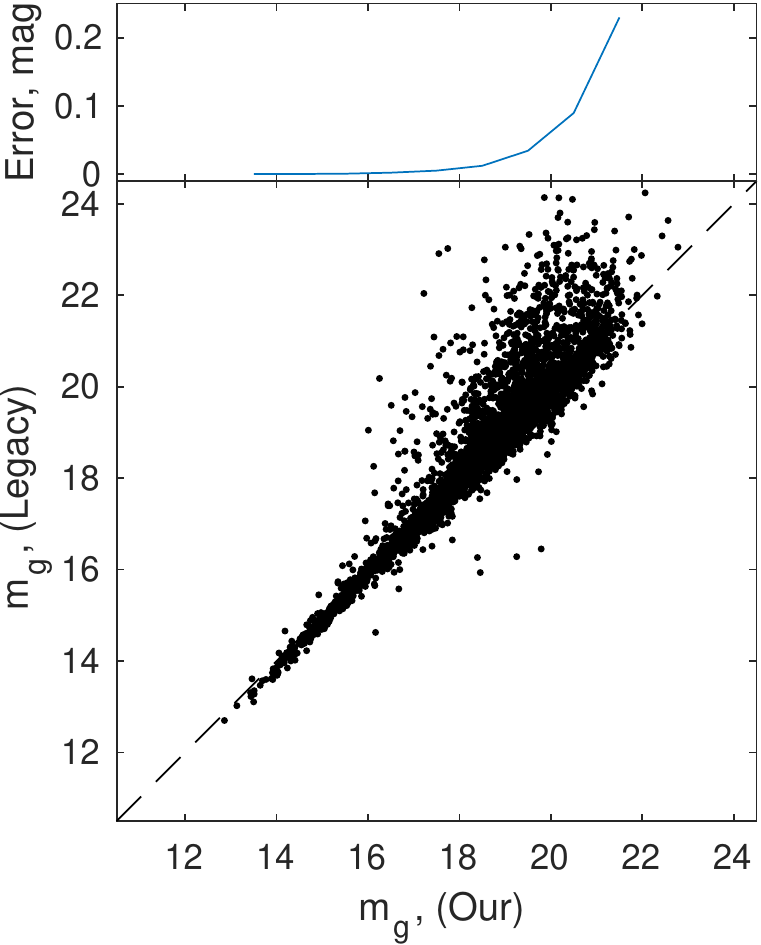}
\caption{Comparison between our $g$-band magnitude and Legacy survey's pipeline based $g$-band magnitude. In the top panel, we show a median error in our $g$-band photometry.}
\label{magcompare}
\end{figure}

\begin{figure}
\label{main}
\includegraphics[width=8cm]{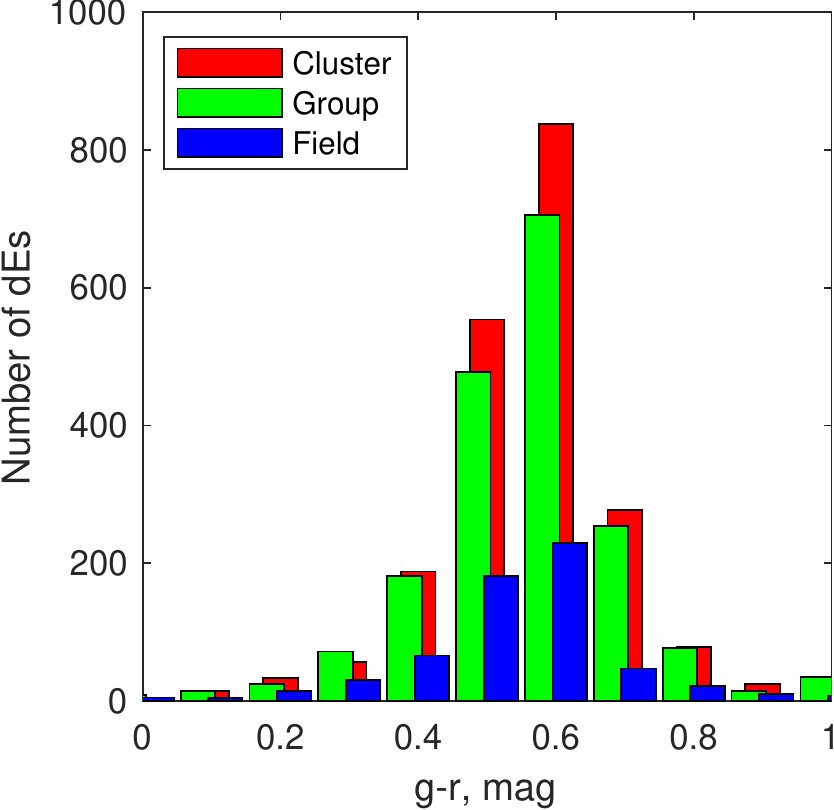}
\caption{The $g-r$ color distribution of our sample dEs. The bin positions of the group and field samples are shifted by +0.01 and $-$0.01 mag, respectively.}
\label{crhist}
\end{figure}

\begin{figure}
\label{main}
\includegraphics[width=8cm]{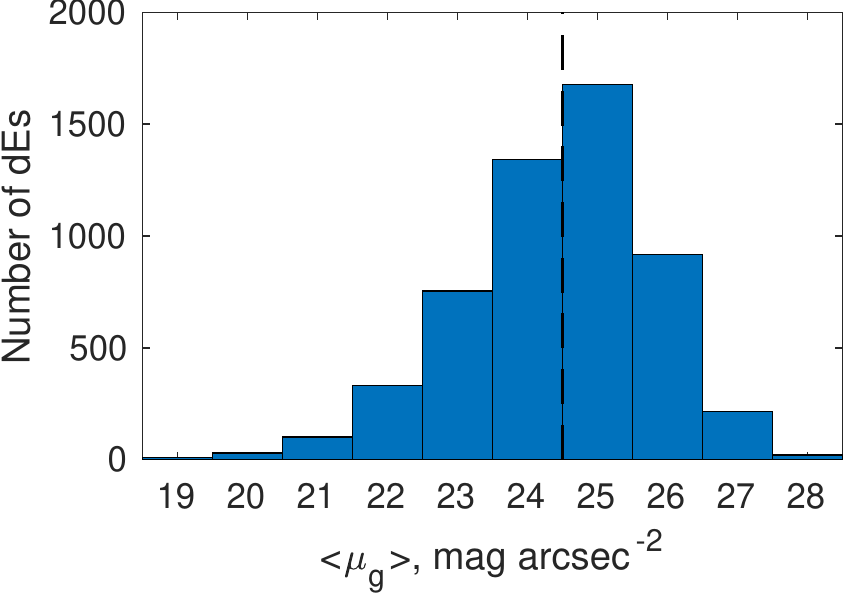}
\caption{The $g$-band mean surface brightness distribution of our sample dEs. The vertical dash line represents the surface brightness limit that is customarily used to define UDGs.}
\label{sfbhist}
\end{figure}

We show the distribution of $g$-band magnitude of our dE sample in Figure \ref{maghist}, which has a peak of around 19 mag. The median magnitude of our sample is 18.68 mag, immediately reflecting that the majority of our dE sample galaxies are fainter than the spectroscopic target selection limit of the SDSS, i.e., 18 mag, as shown by a vertical line. We also overlay the distribution of our spectroscopic sample dE (the blue histogram), and as expected, it mainly overlaps in the brightest part. 

In Figure \ref{crhist}, we show the $g-r$ color distribution of our dE sample, and all three subsample dEs show a single peak Gaussian distribution, with a peak around 0.6 mag. The overall median value of the $g-r$ color of our sample dE is 0.58 mag, with a standard deviation of 0.14 mag.

To compare our measured $g$-band magnitudes with that of the Legacy survey pipeline values, we obtained the Legacy survey magnitudes using the SQL queries within a 6$\arcsec$ radius of the dE center. We used the primary source catalog provided by the tractor. In Figure \ref{magcompare}, we compare our $g$-band magnitudes and the Legacy survey catalog values. We find that these two magnitudes agree well with each other for the bright galaxies, and for the fainter galaxies, it seems that the Legacy tractor pipeline underestimates the flux.

We show the distribution of $\mmu$ in Figure \ref{sfbhist}. It is very clear that most dEs have mean surface brightness higher than 24.5 mag arcsec$^{-2}$, which has been frequently used as a cutoff magnitude to define UDGs, see section \ref{scaling}. Our dE sample has a median value of $\mmu$ = 25.97 mag arcsec$^{-2}$. Although we do not use any magnitude or surface brightness limit for the visual selection while preparing the final catalog, we select dEs brighter than m$_{g}$ = 22.5 mag and  $\mmu$ $<$ 28 mag arcsec$^{-2}$.  We introduced these limits based on the measurement error.


\section{Morphology}\label{morph}

\subsection{Color-magnitude relation}
\begin{figure}
\includegraphics[width=8cm]{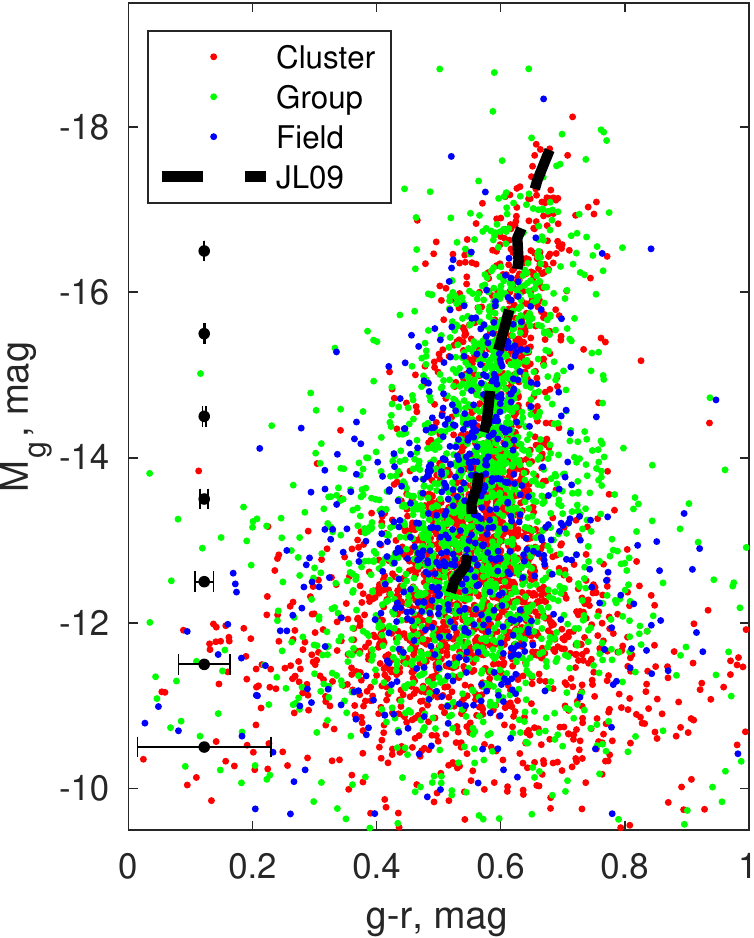}
\caption{The color$-$magnitude relation of our sample dEs. We show cluster, group and field dEs in red, green and blue dots, respectively. The dashed line represents an average color-magnitude measured by \cite{Janz09} for the Virgo cluster dEs using the SDSS data. The SDSS magnitudes are converted to the Legacy DECaLS filter system according to the transformation formula provided by the Legacy web page. We show a typical error bar for each magnitude bin in the black symbol.}
\label{cmr}
\end{figure}

Early-type galaxies follow a well-defined color-magnitude relation (CMR), and it has been shown that they are astonishingly similar for various environments \citep{Visvanathan77,Sandage78}.  However, as noted by \cite{Janz09}, this relation is not linear in going from higher mass elliptical galaxies to lower-mass dEs. In Figure \ref{cmr}, we show the CMR of our dE, and for comparison we overplot an average CMR sequence of Virgo dEs derived by \cite{Janz09}, see black dash line.  Although our dEs are located in diverse environments, they generally follow the CMR defined by Virgo dEs on average. As expected, the scatter at fainter magnitude is larger; in the brighter magnitude ($M_{g}$ $<$ $-$14 mag) regime, however, the CMR is well constrained within the $g-r$ color scattering of 0.2 mag.

\begin{figure}
\includegraphics[width=8cm]{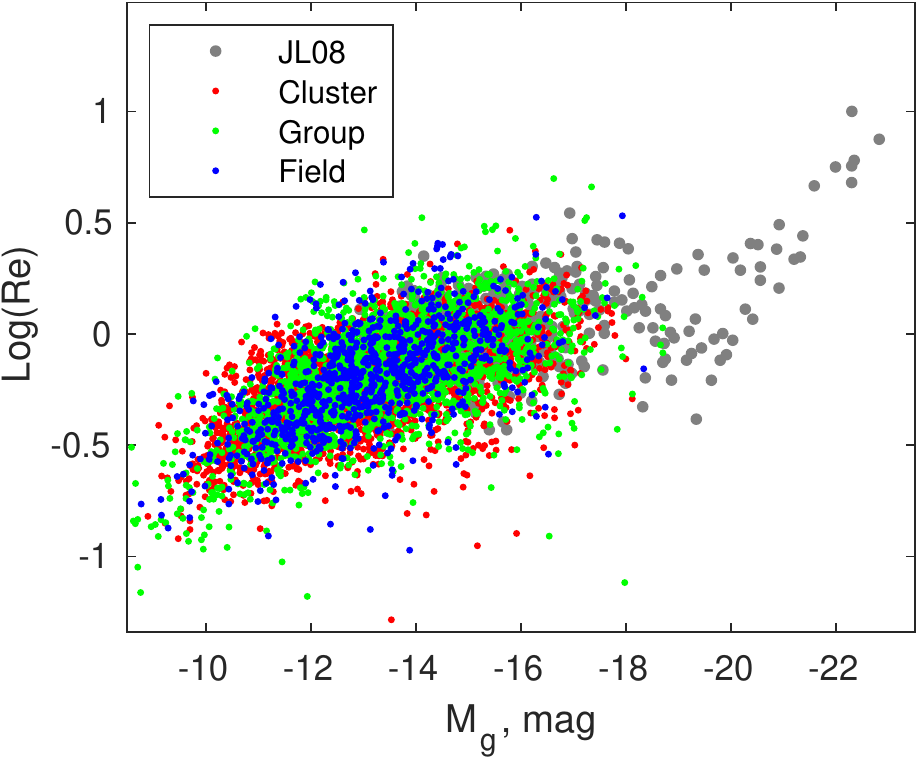}
\caption{The luminosity$-$size relation of early-type galaxies. Data for Es/dEs (gray dots) are from \cite{Janz08}. The symbol color represents cluster, group and field dEs according to Figure \ref{cmr}.}
\label{magsize}
\end{figure}

\begin{figure*}
\label{main}
\includegraphics[width=17cm]{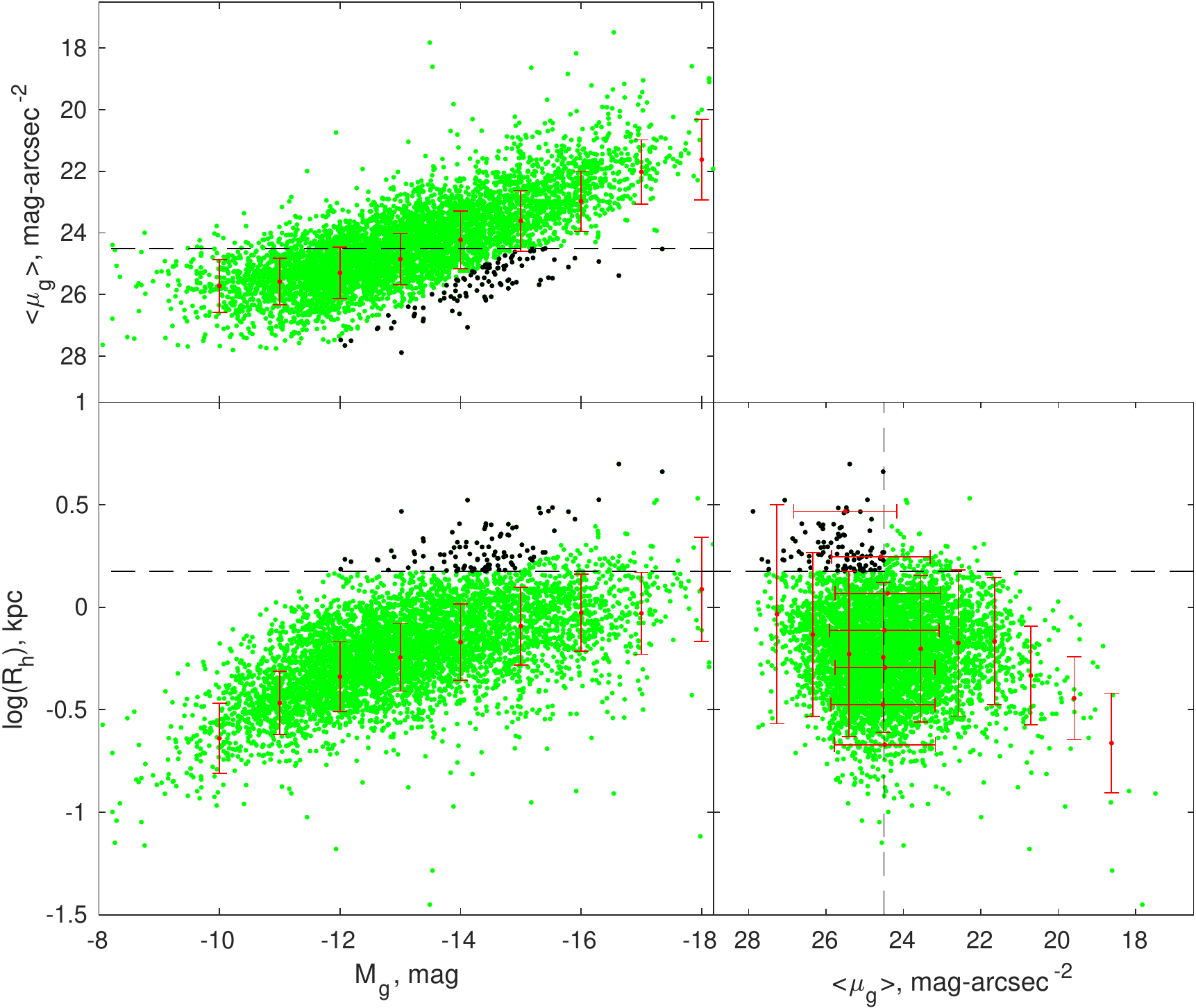}
\caption{The scaling relations between the structural parameters (R$_{h}$ and $\mmu$) and M$_{g}$ of our sample dEs. Green and black dots represent dEs and UDGs, respectively. The threshold on $\mmu$ and R$_{h}$ that are used to define UDGs are shown by the dashed lines. The red dots with a 1-$\sigma$ standard deviation represent the median values of the distribution.}
\label{udgsc}
\end{figure*}

\subsection{Scaling relation and UDG}\label{scaling}

Early-type galaxies obey a scaling relation between their structural parameters (such as $R_{h}$ and $\mmu$) and the total luminosities \citep{Kormendy77,Guzman93}.
In Figure \ref{magsize}, we show the relation between Log($R_{h}$) and the magnitudes for the early-type galaxies.  For comparison, we used a sample of early-type galaxies studied in JL08 \citep{Janz08}, shown by gray color symbols. Our sample dE are shown in red, green, and blue for the cluster, group and field members, respectively. At first glance,  our dE sequence is well matched with the dE sequence of JL08, and as was discussed in  JL08, the dE follows a less steep scaling relation of magnitude and Log($R_{h}$) compared to the scaling relation for Es.

Our sample includes many extended, low-surface brightness galaxies, which satisfy the definition of UDGs, i.e., dEs with $\mmu$ $>$ 24.5 mag arcsec$^{-2}$ and $R_{h}$ $>$ 1.5 kpc \citep{Marleau21}. However, to select UDG there is no standard cutoff limit on $\mmu$. In the literature, we find a range of $\mmu$ from 24 to 25 mag arcsec$^{-2}$ that has been used to select  UDG candidates \citep{Habas20,Lim20}. For this work, we used 24.5 mag arcsec$^{-2}$  in the $g$-band. In our sample, we find 100 dEs that can be classified as  UDGs. Among them, 12 are in the cluster (10 in Virgo and 2 in Fornax), 59 in groups, and 29 in the field environment. On average, we find six UDGs per cluster and 0.22 per group, agreeing with the trend found by  \cite{Burg17}. 

In Figure \ref{udgsc}, we show the interdependence of the structural parameters ($R_{h}$, $\mmu$ and $M_{g}$)  of our sample of dEs and try to explore whether the UDG occupies a special position within it.  In the middle panel, we show the scaling relation between Log($R_{h}$) and $M_{g}$ where the red symbol represents the median value of Log($R_{h}$) in each magnitude bin with a standard deviation shown as an error-bar. On the top panel, we show the relation between $\mmu$ and $M_{g}$, and in the right panel, we show the relation between $\mmu$ and Log($R_{h}$). The UDGs are shown in black symbol and the green points represent dEs.

\begin{figure*}
\label{main}
\includegraphics[width=18cm]{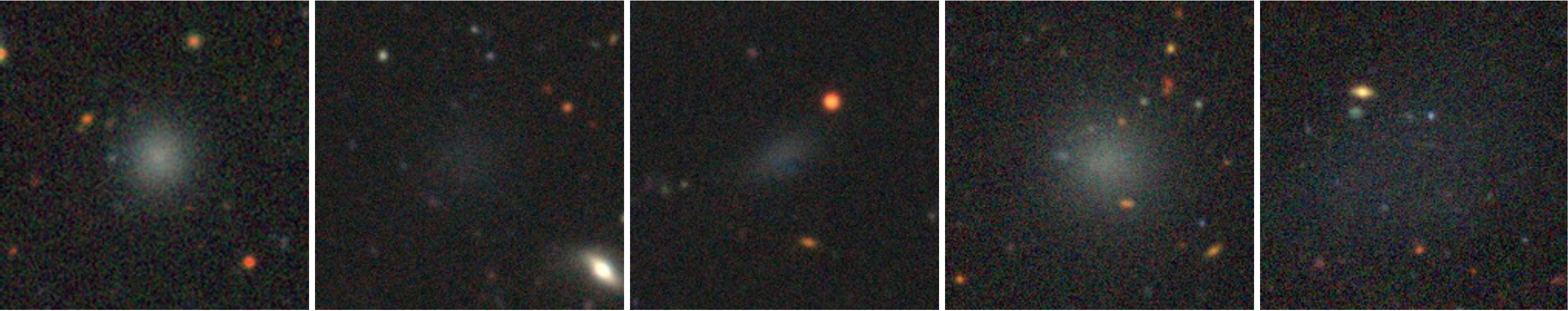}
\includegraphics[width=18cm]{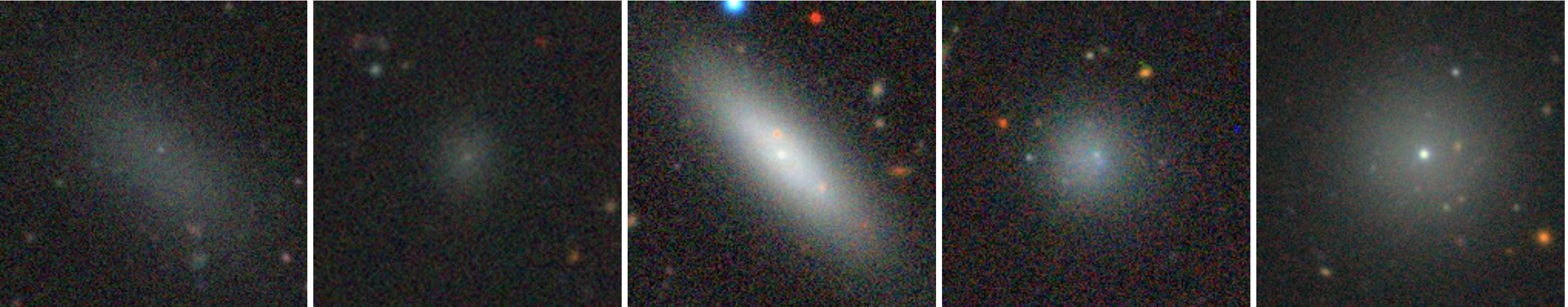}
\caption{Examples of nonnucleated (upper row) and nucleated dEs (lower). All stamp images have FOV of 1\arcmin$\times$1\arcmin.}
\label{nucexample}
\end{figure*}

\begin{figure}
\label{main}
\includegraphics[width=8cm]{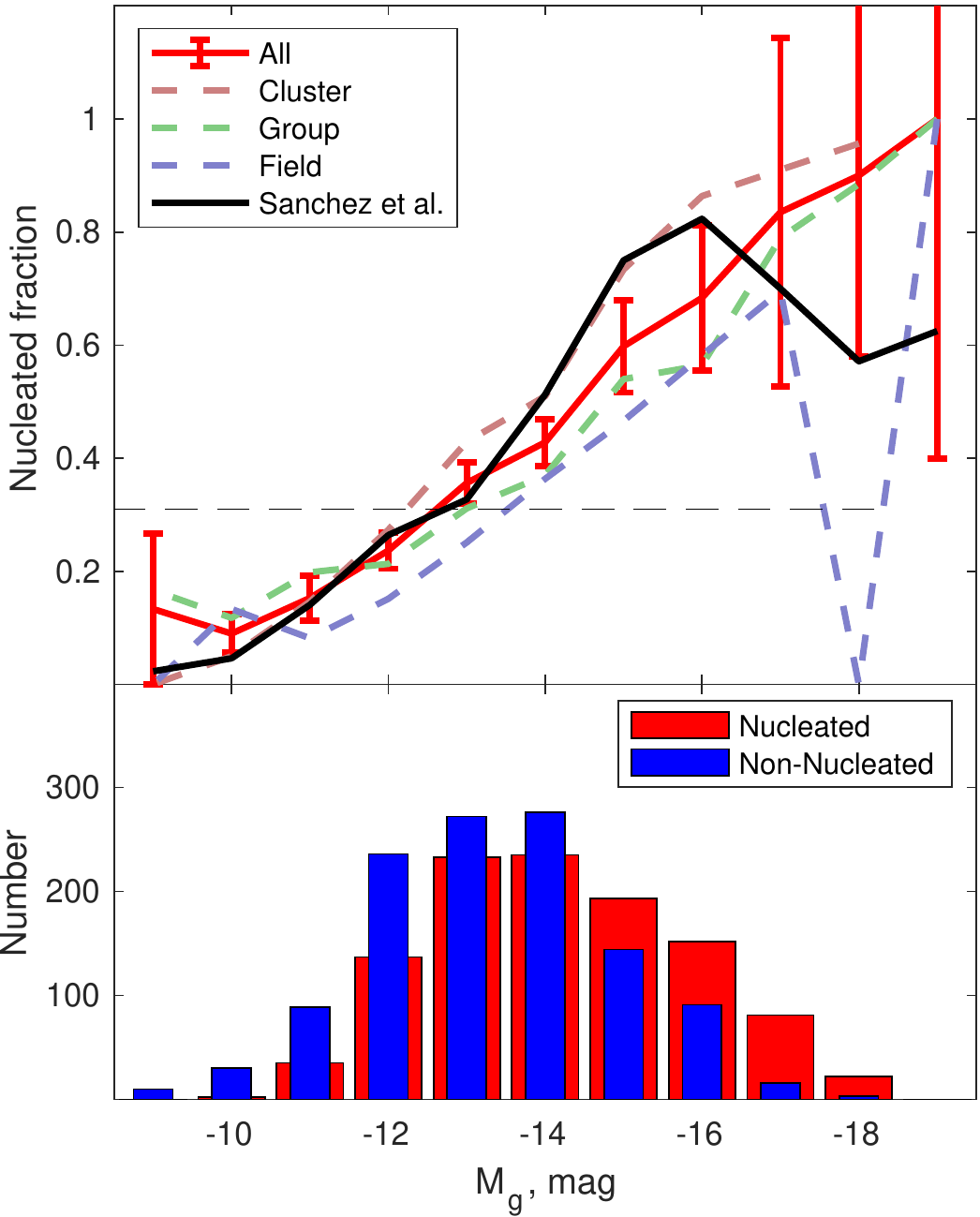}
\caption{Top: the nucleated dwarf fraction with respect to magnitude. The black line represents the nucleated dE fraction measured by \citet{Sanchez19} for the Virgo Cluster, and the red line represents a nucleated fraction of our entire dE population. The nucleated fractions of dEs for clusters, groups and fields are shown with magenta, green, and blue lines, respectively.\\
Bottom: magnitude distribution of nucleated (red histogram) and nonnucleated (blue) dEs. }
\label{nufrac}
\end{figure}

 \begin{figure*}
\label{main}
\includegraphics[width=18cm]{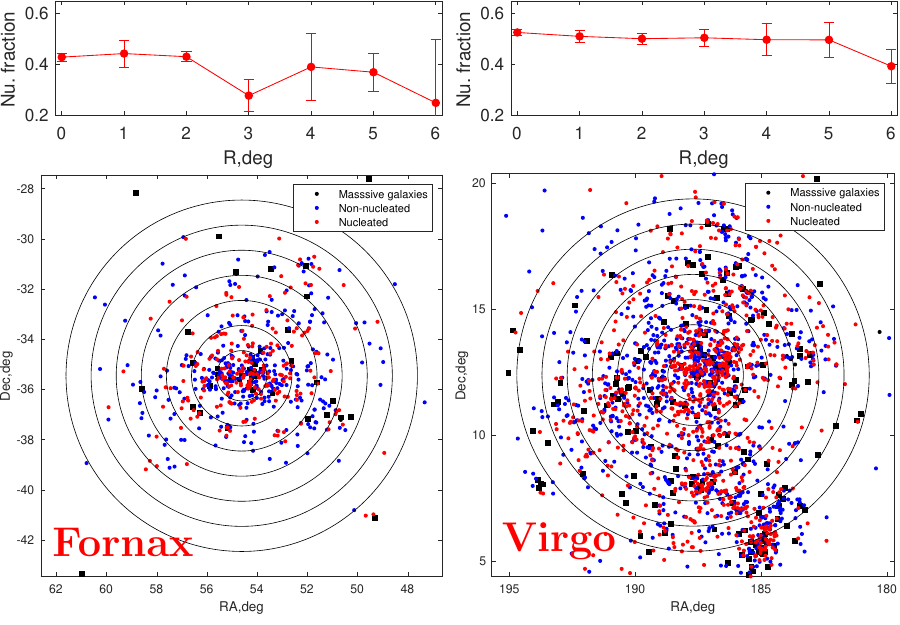}
\caption{Distribution of nucleated (red dots) and nonnucleated dEs (blue dots) inside Virgo and Fornax clusters. Fornax is shown on the left, centering at NGC 1407, and Virgo is shown on the right, centering at M87. The position of bright member galaxies of clusters are shown by black squares. The black concentric rings represent the annular aperture of one-degree width, where we measure the nucleated fraction. The results are shown in the top panels. }
\label{nucdis}
\end{figure*}

As per the definition, UDGs occupy the upper left corner of the $\mmu$ and Log($R_{h}$) relation defined by the crossing demarcation lines. Given the large scatter of the data points, we find no special clustering of UDGs distinct  from the overall dE population. They are, however, a one-sigma outlier in the size-magnitude relation. The middle panel reveals that the limit of $R_{h}$ $>$ 1.5 kpc may well separate UDGs from dEs in the fainter regime but not in the brighter regime as we can see that all dE fainter than $-$15.96 mag with $R_{h}$ larger than 1.5 kpc can be classified as UDGs. On the other hand, the top panel reveals that the demarcation line of $\mmu$ = 24.5 mag arcsec$^{-2}$ cannot uniquely separate UDGs from dEs, and many dEs have mean surface brightness higher than $\mmu$ = 24.5 mag arcsec$^{-2}$, in fact, the majority our sample dEs have $\mmu$ $>$ 24.5 mag arcsec$^{-2}$.

Our analysis reinforces the conclusion noted by \cite{Conselice18,Danieli19}, that UDGs do not occupy a special position in the parameter space of early-type galaxies, defining them as a special class of objects, nor are their structural parameters uniquely different from the overall dE population, being only a 1$\sigma$ outlier on the Log($R_{h}$) and $M_{g}$ relation.   We show that the UDGs and dEs have a significant overlap in parameter space that supports a conclusion that the dEs and UDGs are essentially the same objects, where the latter represents an extended subpopulation of the former.

\subsection{Presence of Nucleus}
There are a plethora of works on nuclei of dEs in the cluster environments, particularly Virgo, Fornax, and Coma \citep{Cote06,Paudel11,Brok14,Ordene18,Wittmann19,Sanchez19,Poulain21}. In this work, we homogeneously investigate the presence of nuclei in dEs located in diverse environments, i.e., cluster, field, and group. Although there is no standard definition of the nucleus in dE, the nucleus is usually defined as the existence of a luminosity excess over the main stellar distribution in the core region \citep{Sanchez19,Paudel20}. In this work, we consider the presence of a compact point source of point-spread function (PSF) size at the core region as a nucleus. To identify a nucleus at the center of a dE, we carefully examine the color image cutout of every dE with an FOV of 1\arcmin$\times$1\arcmin. During the visual examination, we excluded some dEs with central star-formation and dEs with high central surface brightness. Because the central star-forming dEs have irregular blue core making confusion in the identification of the nucleus \citep{Lisker06,Urich17,Paudel20}. High central surface brightness dEs are primarily compact objects. It is impossible to visually identify the presence of a separate PSF component at the center of these compact galaxies. In any case, cE, M32-like galaxies, rarely host a central nucleus.

We can classify 4767 dEs as nucleated or non-nucleated, and among them, 2065 dEs are nucleated. A few randomly selected example dEs of both nucleated and nonnucleated classes are shown in Figure \ref{nucexample}, where the top panel represents nonnucleated and the bottom panel represents nucleated dEs.

In Figure \ref{nufrac}, we show the magnitude distribution of nucleated and nonnucleated dEs. The magnitude distribution shows that the nuclei are common in bright dEs, and they become rare in the least luminous dEs, as we can see that the red histogram dominates the brighter part while the blue histogram dominates the faint end. We show the relation between nucleated fraction and magnitude in the top panel. The red line represents the nucleated fraction of the overall dE population, and magenta, green, and blue lines represent the nucleated fraction of cluster, group, and field subsampled dEs, respectively. The error bar denotes the fraction of dEs that we were not able to classify into nucleated and non-nucleated. For comparison, we also show the measurement of the nucleated fraction by \citep{Sanchez19} for the Virgo Cluster core-region of 4 ($2\degr\times2\degr$) square degrees. Our overall nucleated fraction reasonably agrees with \citep{Sanchez19} result. A careful inspection of these results shows that the cluster environment is likely to have a higher nucleated fraction than the group. The field environment hosts a comparatively lower nucleated fraction than both clusters and groups. However, these differences are well within the measurement uncertainty.

We study a radial dependence of nucleated fractions inside the Virgo and Fornax cluster, where the nucleated fraction are binned in the annular aperture of 1$\degr$ width, see bottom panel of Figure \ref{nucdis}. We show the result in the top panel and find that the nucleated fraction is mildly anticorrelates with cluster-centric distance. Comparatively, the Fornax Cluster shows slightly steeper anticorrelation than the Virgo Cluster. Within the central 1$\degr$ radius area, Virgo and Fornax host nucleated fractions 43\% and 52\%, respectively. At the outer region, beyond a 5$\degr$ cluster-centric distance, the nucleated fraction reaches down to 30\% for both clusters.

\subsection{Morphological Feature}

\begin{figure}[h]
\label{main}
\includegraphics[width=8.5cm]{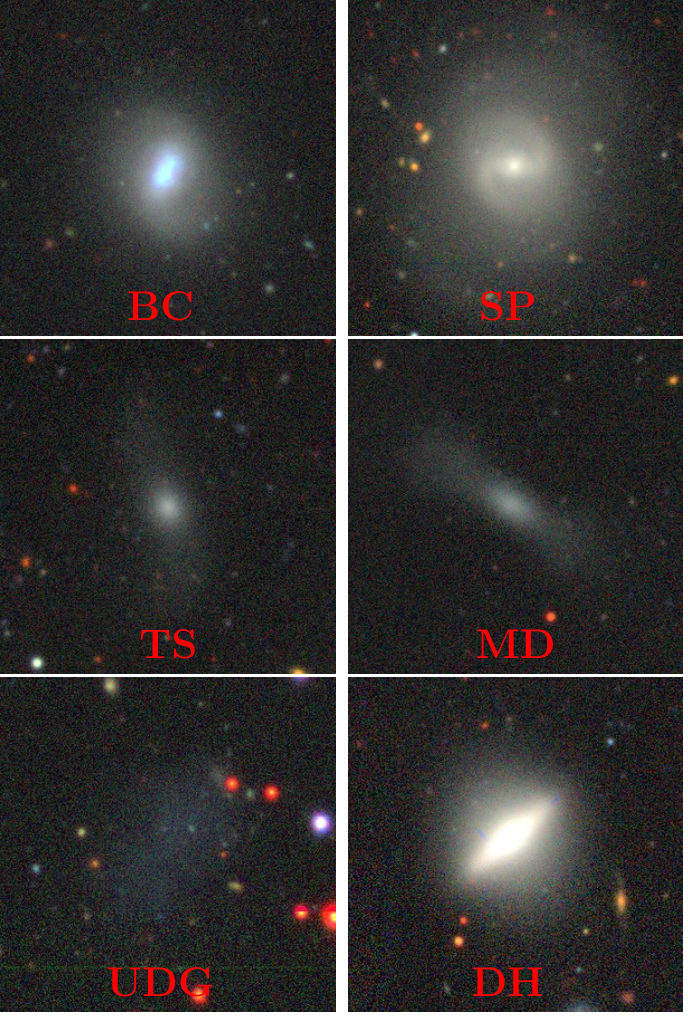}
\caption{Examplary dEs of the different classes of dE according to their morphological feature and color (see text for detail).}
\label{dEclass}
\end{figure}

We carefully examine dE images and classify them according to their low-surface brightness and color characteristics. The primary purpose of this classification scheme is to recognize some unique dEs that are not frequent in observation. Indeed, we excluded most of the normal dEs in this classification, which host no special feature. Using a broad scheme of classifiers, we try to identify six distinctive morphological feature present in dE. Indeed, these features are already well investigated in some individual cases of dEs, particularly in the cluster environment. For example, the presence of spiral arm and blue center in dEs were studied by \cite{Lisker06,Lisker06a} and tidal features, like a shell or the stellar, steams, were studied in \cite{Paudel14a,Paudel17a}. This sample aims to provide homogeneous statistics of these features' presence in dEs in various environments. Below, we itemize these six general categories. 

\begin{itemize}
\item Blue core (BC): the presence of blue core region as a sign of recent star-forming activity \citep{Lisker06}.
\item Spiral arm (SP): the presence of spiral arm or bar commonly called disk feature dEs \citep{Lisker07,Smith21}.
\item Tidal stream (TS): tidally stretched dE due to interaction with nearby massive galaxy \citep{Paudel14a}.
\item Merging dwarf (MD): the low-surface brightness features that are most likely to be originated through the dwarf-dwarf merger, e.g., shell or tidal tail \citep{paudel14b,Paudel17a,Paudel18}.
\item UDG: ultra-diffuse galaxies, see section \ref{scaling}
\item Disk-halo (DH): we identify some dEs that possess a prominent edge-on disk with a rounder and diffuse stellar halo. 
\item Normal dEs(ND): rest of dEs, which is not classified by any of the above classifiers. 
\end{itemize}

We show examples of these subclasses in Figure \ref{dEclass}. We find that 288 dEs possess a central blue core that may reveal recent star-forming activity at the center \citep{Urich17}. These blue core dEs are mostly located at the outskirts of cluster or group environments. They could instead represent transition-type dwarf galaxies that are in the process of being transformed into the dEs after cessation of star-formation activity \citep{Koleva13}. 

\begin{figure}
\includegraphics[width=8.5cm]{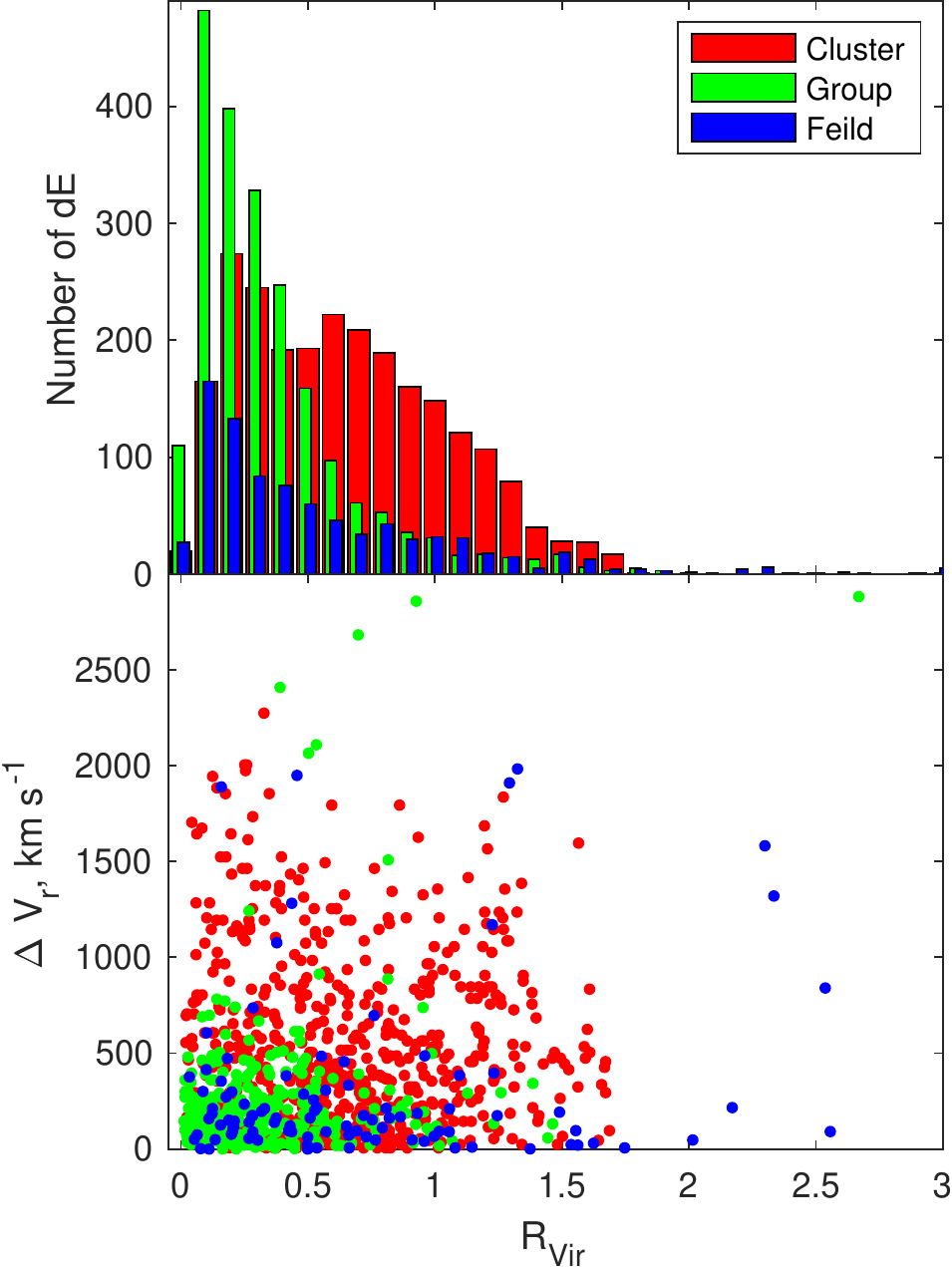}
\caption{(Top) The distribution of sky-projected separation of dEs from the host center. The x-axis is normalized by the assumed viral radius of each cluster, group and field. Red, blue, and green histograms are for clusters, groups and fields, respectively. (Bottom) Phase space distribution of our spectroscopic dE sample. Red, blue, and green dots are for clusters, groups and fields, respectively.}
\label{dishist}
\end{figure}

We find that 17 dEs possess prominent spiral arms. However, note that this number could be higher. Because we merely select them after visual assessment and no unsharp masking has been used to identify these features as in \cite{Lisker07}. As \cite{Michea21} have shown that we may need to use a more sophisticated image analysis procedure to reveal the disk feature hidden inside the smooth stellar halo of dEs. 
 
Approximately 1.3\% (71 out of 5405) of dEs are experiencing tidal distortion due to the effect of the tidal potential of nearby massive galaxies. Out of 71 tidally disturbed dEs, only 13 are located inside the cluster environment, and an overwhelming majority (58 out of 71) of them are located in the group or field environment. This result, in fact, agrees with \cite{Paudel14a} study, which shows that the tidal harassment process is more frequent in a low-dense environment like a group and field compared to a dense cluster. Only 0.3\% of dEs show tidal features that may have originated from merging even smaller dwarf galaxies.

In this work, we introduce a unique class of dEs, for the first time, which possesses two components morphological features, an edge-on disk with a round stellar halo. These two-component systems may be a scaled-down version of the disk-halo system (DH) of our galaxy Milky-Way. These DH dEs could be an edge-on view of spiral arm dE viewed on the higher inclination angle of the disk, where the diffuse halo remains hidden behind the prominent spiral arm \citep{Smith21}. In that case, dE (SP) and dE (DH) could represent the same class but are different in viewing angle. Nevertheless, we do not see the underlying spiral arm due to high inclination, and we keep them in a separate class. \trev{ To get a statistical estimate, we calculate the probability of an edge-on view of a spiral-arm galaxy at a random viewing angle. Using an assumption of a thin disk \cite[intrinsic thickness q$_{0}$=0.2,][]{Janssen10}, we find that there is $\approx$30\% chance that the thin disk can be viewed as edge-on (b/a $<$ 0.5). We have found only five DH dEs, which is 29\% SP dEs. }

\section{Environment}\label{env}
 
 \begin{figure}
\includegraphics[width=8.5cm]{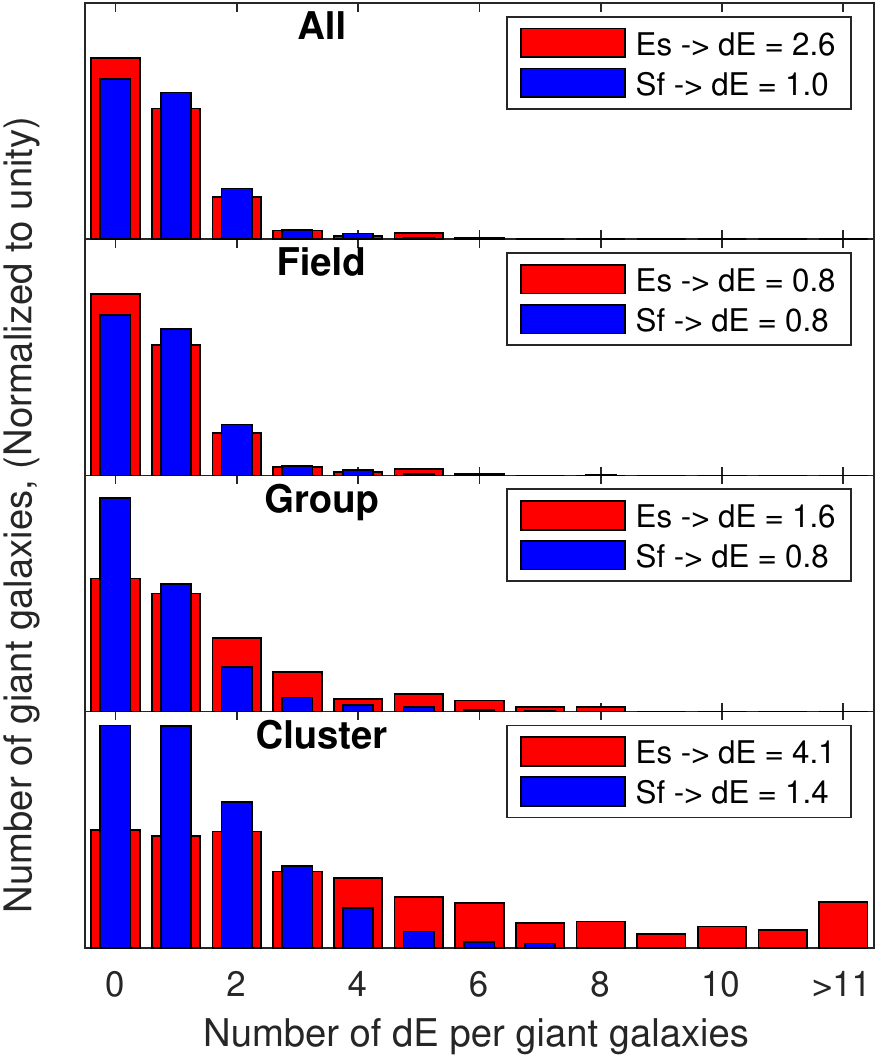}
\caption{Frequency of bright galaxies around different types of dEs. The first to last rows are for all, field, group and cluster dEs. The blue (red) histograms represent the number of Sfs (Es) around dEs within the 300 kpc sky-projected radius.}
\label{edensity}
\end{figure}

As per our selection procedure, our dEs are either inside the cluster/group or near bright field galaxies (hereafter the host). In this subsection, we explore the dE positions with respect to the host and morphology types of their companion bright galaxies (BGs, $M_{k}$ $<$ $-$21 mag).

In the bottom panel of Figure \ref{dishist}, we show a phase-space position of our dEs within their respective host environment. In the top panel, we show the distribution of sky-projected separation of the dEs from the host center, where the host-centric distance is normalized by viral radii of their respective cluster, group, and field. We used 1.7 and 1.2 Mpc for Virgo and Fornax Clusters, respectively. For the group, we used 800 pc, and for the field we used 300 pc.

Almost all dEs are located well within the 1.5 virial radii; however, a few field dEs are significantly away from the host. Although they are few, we noticed that these field outliers are a spectroscopically selected sample of dEs from the SDSS. We have used an all-sky catalog from the SDSS rather than searching around predefined areas of the cluster, group, or filed. 

In terms of number count, the two clusters contribute nearly half, 2437 (1837 and 864 in Virgo and Fornax, respectively) out of 5405. We find that there are 7.9 dEs per group and 1.5 dEs per field, on average. We noticed that there are many fields which contain no dEs. Out of total of 586 fields we have inspected, only 233 fields host one or more dEs. We further classify the field host as an early-type (E) and star-forming (Sf) and find that there are 437 Sps and 153 Es field hosts. We find that out of 437 fields with a Sfs host, only 154 have one or more dEs, and out of 153 field with an Es host, only 79 have one or more dEs, which indicates a higher chance of hosting a dE by an Es than Sfs. On average, we find that early-type fields host a significantly higher (2.8) dEs per field compared to star-forming field hosts, which have 1.0 dE per field, on average.

Further, we explore the relationship between the location of dEs and morphological types of neighbor BGs, regardless of their environment. We counted BG frequency of each type around each dE within a 300 kpc radius. The result of this exercise is shown in Figure \ref{edensity}. We find a higher frequency of dEs around Es than Sfs. On average dEs have a 2.5 times higher chance to have an Es as a neighbor than Sfs.

Indeed, this result might have been heavily influenced by the cluster environment, where E dominates by number. We performed this exercise again, separating the cluster, group, and field. We find a similar trend in the cluster and group dEs, but in the field environment, where Sf dominates, we find an equal chance to have both E and Sf as a neighbor of dE on average. The results are listed in Figure \ref{edensity}.

\section{Conclusion and Discussions}\label{conc}

We have identified 5405 dEs located in the various environments, i.e., cluster, group, and field, spanning the luminosity range $-$18 $<$ $M_{g}$ $<$ $-$8 mag. They are selected via visual inspection of 7643 deg$^{2}$ area of the Legacy survey $g-r-z$ combined tri-color images, which include two clusters (Virgo and Fornax), 265 groups, and around 586 field galaxies. Our dE catalog provides several different metrics by which we quantify morphological and environmental properties. The salient feature of this catalog lies in their morphological properties, as we identified morphological characters such as nucleated, tidal, and UDGs.

 \begin{figure}
\includegraphics[width=8.5cm]{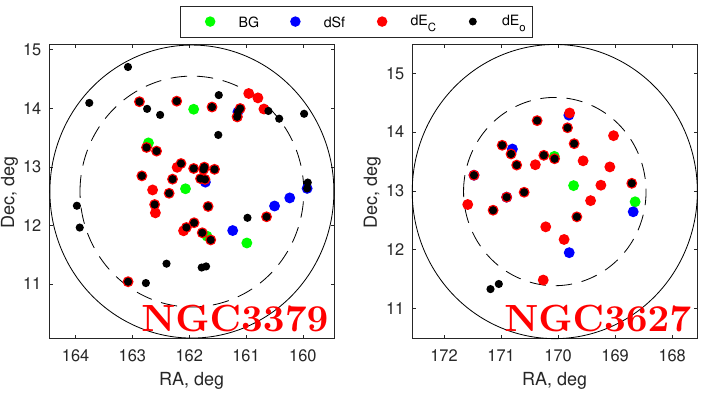}
\caption{We compare dEs detection in two groups NGC\,3379 and NGC\,3627, with detection of  C22, where the circle represents our search area. Giant galaxies are shown in green symbol. C22 identified, star-forming (dSf) and non-star-forming (dE$_{C}$) dwarf galaxies are shown in blue and red, respectively. Our identified dEs (dE$_{o}$) are shown in black. The solid and dashed circle represent our and C22 search areas, respectively. }
\label{crcomp}
\end{figure}

Given the nature of the selection procedure, it is not trivial to calculate the completeness limit of our sample. Although we aim to search dEs of up to $z$ = 0.01, we should emphasize that this is not a volume-limited sample in many regards. Indeed, we have not explored all sky areas of this volume. In addition, one of the main caveats of this work is the distance of candidate dEs, and in most cases we do not have that information. There may be a faction of dEs that are foreground or background galaxies or at least not directly related to the host. To confirm their distance and association with the host environment, we need radial velocity information, which we do not have in most cases.

This work is explicitly designed to find dEs in an extensive database, and there are not many surveys of this kind for which we can make a direct comparison.  We found that \citeauthor{Carlsten22} (\citeyear[][hereafter C22]{Carlsten22}) searched satellite dwarf galaxies (of all types) in the Local Volume  ($D$ $<$ 12 Mpc) in the Legacy survey imaging.

Unfortunately, we found that only two areas around NGC\,3379 group and NGC\,3627 group broadly overlap with our searched sky regions. In the case of NGC\,3379, they have found 36 dwarf galaxies, and among them, 33 can be classified as dE from our visual inspection; see Figure \ref{crcomp}. Our catalog recovers 26 (78\%) out of 33. Similarly, we recovered 16 (64\%) out of 25 dEs in the case of NGC\,3627. We noticed that those unrecovered 16 (9+7) dEs are mostly fainter dEs of $m_{g}$  $\gtrsim$ 20 mag. In the C22 search area, we identified 8 additional dEs, bringing the total number of identified dEs in two systems to 50. Calculating the recovery fraction of the C22 dE sample, we have recovered 42 (72\%) dEs, which is smaller compared to the overall recovery fraction for $m_{g}$ $<$ 20 mag (i.e., 87\%). Multiple factors may have contributed to this discrepancy, including environmental differences. The cluster environment plays a crucial role in shaping the brightness distribution of dEs, where brighter dEs are more likely to be formed \citep{Geha12} and they have higher recovery fractions. In this sample, a substantial portion of dEs are found within the cluster environment. Also, a potential memory bias in our calculation may have contributed to the high recovery fraction.

Based on this extensive data set, our results are summarized as follows:

\begin{enumerate}
\item  Systematic analysis of the derived structure parameters of dEs reveals that the dEs generally follow a universal relation between magnitude, size, and mean surface brightness regardless of their host environment. 
\item Using a standard definition, we identify 100 dEs in our sample that can be classified as UDGs. However, we find that UDGs do not occupy a special position in the parameter space of dEs to define them as a special class of objects. In addition, their structural parameters are not uniquely different from the overall dE population, being only one sigma outlying from the Log($R_{h}$) and $M_{g}$ relationship. 
\item We identify that only 40\% of our sample dEs host a central nucleus, and among the UDG population, the vast majority, 79 out of 100, are non-nucleated. We find a mild radial dependence of nucleated fraction in both the Virgo and Fornax Clusters, and overall, the nucleated fraction declines as we go from the higher-density environment cluster to the low-density environment.
\item We find that about 1.3\%of dEs suffer a tidal disturbance from nearby massive galaxies and only 0.03\%suffer recent dwarf-dwarf merging.
\item  We find that dEs are found more frequently in environments where massive galaxies are already quenched.

\end{enumerate}

We hope that this catalog will serve as a valuable resource for follow-up studies on dEs. In particular, considering the laborious nature of visual inspection and classification, this data set will be useful for training and testing machine learning approaches. In future work, we plan to test the ability of convolutional neural networks to detect and characterize dEs with a view to applying them to a larger set of imaging data.

\newpage
\begin{acknowledgments}
S.P. and S.-J.Y., respectively, acknowledge support from the New Researcher Program (Shinjin grant No. 2019R1C1C1009600) and the Midcareer Researcher Program (No. 2019R1A2C3006242) through the National Research Foundation of Korea.

This study is based on archival images and spectra from the Sloan Digital Sky Survey and images of Legacy survey. The full acknowledgment for the SDSS can be found at http://www.sdss.org/collaboration/credits.html.  Funding for the SDSS has been provided by the Alfred P. Sloan Foundation, the Participating Institutions, the National Science Foundation, the U.S. Department of Energy, the National Aeronautics and Space Administration, the Japanese Monbukagakusho, the Max Planck Society, and the Higher Education Funding Council for England. The SDSS website is http://www.sdss.org/. 

The Legacy Surveys consist of three individual and complementary projects: the Dark Energy Camera Legacy Survey (DECaLS; NOAO Proposal ID \# 2014B-0404; PIs: David Schlegel and Arjun Dey), the Beijing-Arizona Sky Survey (BASS; NOAO Proposal ID \# 2015A-0801; PIs: Zhou Xu and Xiaohui Fan), and the Mayall $z$-band Legacy Survey (MzLS; NOAO Proposal ID \# 2016A-0453; PI: Arjun Dey). DECaLS, BASS, and MzLS together include data obtained, respectively, at the Blanco telescope, Cerro Tololo Inter-American Observatory, National Optical Astronomy Observatory (NOAO); the Bok telescope, Steward Observatory, University of Arizona; and the Mayall telescope, Kitt Peak National Observatory, NOAO. The Legacy Surveys project is honored to be permitted to conduct astronomical research on Iolkam Du'ag (Kitt Peak), a mountain with particular significance to the Tohono O'odham Nation. The full acknowledgment for the Legacy Surveys can be found at https://www.legacysurvey.org/acknowledgment
\end{acknowledgments}


\newpage
\begin{table}
\centering
\caption{Catlog table (a representative sample). }
\begin{tabular}{cccccccccccc}
\hline
 ID & R.A. & Decl.   & m$_{g}$ & M$_{g}$ &     $\mmu$ & R$_{h}$  & z & Nuc & Class & Env.type & Host name \\
 (hhmmssddmmss) & (deg) & (deg) & (mag) &  (mag) & (mag arcsec$^{-2})$ & (kpc) &  & & & & \\
 (1) & (2 &   (3) &  (4) &  (5) &  (6) &  (7) &  (8) &  (9) &  (10) &  (11) &  (12) \\
 \hline
000033+165423  & 000.1392  & 16.9065  & 18.58  & -12.75  & 24.58  & 0.52(06.33)  & ------  & 0  & ND  & 2  & 218\\
000044+152725  & 000.1855  & 15.4572  & 19.52  & -11.81  & 26.10  & 0.68(08.25)  & ------  & 0  & ND  & 2  & 218\\
000222+160140  & 000.5929  & 16.0279  & 19.56  & -11.77  & 24.16  & 0.27(03.31)  & ------  & 0  & ND  & 2  & 218\\
000248+163546  & 000.7041  & 16.5962  & 19.21  & -12.12  & 25.41  & 0.57(06.94)  & ------  & 1  & ND  & 2  & 218\\
000250+163742  & 000.7105  & 16.6284  & 19.38  & -11.94  & 26.52  & 0.88(10.67)  & ------  & 1  & ND  & 2  & 218\\
000306+161829  & 000.7791  & 16.3083  & 18.27  & -13.06  & 26.30  & 1.34(16.13)  & ------  & 0  & ND  & 2  & 218\\
000324+161111  & 000.8508  & 16.1865  & 18.54  & -12.79  & 25.79  & 0.93(11.23)  & ------  & 0  & ND  & 2  & 218\\
000417+204430  & 001.0715  & 20.7419  & 18.48  & -14.50  & 24.24  & 0.99(05.66)  & ------  & 1  & ND  & 3  & 279\\
000844+143529  & 002.1856  & 14.5916  & 16.65  & -14.68  & 24.04  & 0.99(11.97)  & ------  & 1  & ND  & 2  & 218\\
001746+113147  & 004.4420  & 11.5299  & 20.37  & -11.45  & 26.57  & 0.65(06.95)  & ------  & 0  & ND  & 3  & 1160\\
002922-332302  & 007.3456  & -33.3839  & 21.22  & -10.07  & 26.47  & 0.38(04.48)  & ------  & 0  & ND  & 2  & 1851\\
003102-331613  & 007.7622  & -33.2705  & 15.97  & -15.32  & 23.68  & 1.18(13.95)  & ------  & 1  & SP  & 2  & 1851\\
003155-331600  & 007.9815  & -33.2668  & 16.65  & -14.64  & 24.28  & 1.14(13.41)  & ------  & 0  & TS  & 2  & 1851\\
003258-324548  & 008.2426  & -32.7636  & 18.15  & -13.14  & 24.61  & 0.66(07.82)  & ------  & 1  & ND  & 2  & 1851\\
003351-275024  & 008.4651  & -27.8401  & 17.33  & -14.52  & 25.77  & 2.16(19.46)  & ------  & 0  & UD  & 3  & 2052\\
003414-081008  & 008.5585  & -8.1691  & 16.66  & -15.52  & 23.29  & 1.04(08.48)  & ------  & 0  & ND  & 3  & 2081\\
003506-280306  & 008.7789  & -28.0518  & 17.53  & -14.31  & 25.11  & 1.45(13.09)  & ------  & 0  & ND  & 3  & 2052\\
003637-082457  & 009.1542  & -8.4160  & 20.12  & -12.06  & 25.94  & 0.71(05.82)  & ------  & 0  & ND  & 3  & 2081\\
004103-210808  & 010.2636  & -21.1356  & 19.49  & -12.38  & 24.17  & 0.38(03.44)  & ------  & 1  & ND  & 3  & 2478\\
004434-314204  & 011.1453  & -31.7013  & 21.12  & -10.51  & 24.98  & 0.23(02.36)  & ------  & 0  & ND  & 2  & 2778\\
004643-313058  & 011.6808  & -31.5163  & 19.20  & -12.44  & 25.14  & 0.61(06.15)  & ------  & 0  & ND  & 2  & 2778\\
005124-065818  & 012.8526  & -6.9719  & 20.90  & -11.38  & 25.65  & 0.45(03.55)  & ------  & 0  & ND  & 3  & 2980\\
005236-310921  & 013.1524  & -31.1561  & 16.15  & -15.22  & 22.68  & 0.71(08.09)  & ------  & 0  & BC  & 2  & 3089\\
005249-305753  & 013.2051  & -30.9648  & 20.11  & -11.25  & 24.95  & 0.32(03.70)  & ------  & 0  & ND  & 2  & 3089\\
005733-040804  & 014.3901  & -4.1345  & 19.64  & -13.43  & 25.23  & 0.97(05.25)  & ------  & 0  & ND  & 3  & 3429\\
005914-075715  & 014.8103  & -7.9544  & 19.73  & -12.63  & 25.49  & 0.68(05.67)  & ------  & 0  & ND  & 3  & 3572\\
005916-073511  & 014.8192  & -7.5867  & 19.92  & -12.44  & 24.77  & 0.45(03.72)  & ------  & 1  & ND  & 3  & 3572\\
005949-071341  & 014.9559  & -7.2281  & 20.37  & -11.99  & 25.37  & 0.48(03.99)  & ------  & 1  & ND  & 3  & 3572\\
010149-073750  & 015.4583  & -7.6308  & 18.69  & -13.67  & 25.45  & 1.08(08.95)  & ------  & 1  & ND  & 3  & 3572\\
010339-061053  & 015.9145  & -6.1814  & 18.81  & -14.35  & 24.99  & 1.18(06.85)  & ------  & 1  & ND  & 3  & 3768\\

\hline
\end{tabular}
\\

Note: Column (1): ID in $hhmmssddmmss$ format. Column (2): R.A. Column (3): Decl. Column (4): g-band magnitude. Column (5): g-band absolute magnitude. Column (6): mean surface brightness. Column (7): Half-light radius in kpc (in arcsec). Column (8): z. Column (9): nucleated or not $-$1 for yes and 0 for no. Column (10): dE morphlogical sub-class $-$see Section 4.4 Column (11): Environment $-$1 for cluster, 2 for group and 3 for field. Column (12): Host environment name given in PGC number. 
 (This table is available in its entirety in machine-readable form.)
\label{catlog} 
\end{table}

\end{document}